\documentstyle[prl,aps,preprint]{revtex}
\def \sin {{\rm sin}}
\def \cos {\ {\rm cos}}

\def \non{\nonumber}
\def \wzw {$SL(2,R)\times SU(2)\ $}

\def \de {\del}
\def \A {{\cal A}}
\def \AA {{ \cal B}}

\newcommand{\rf}[1]{(\ref{#1})}

\def \foot {\footnote}
\def \bi{\bibitem}
\def \la {\label}

\def \ha {{1 \over 2}}

\def \del{\partial}

\def \ci {\cite}

\def \A {{\cal A}}
\def \AA {\td {\cal A}}

\def\vp{{\bf p}}

\def \bi{\bibitem}

\def \foot{\footnote}
\def\be{\begin{equation}}
\def\ee{\end{equation}}

\def \k {\kappa}

\def \del {\partial}
\def \bd {\bar \partial }

\def \ha{{\textstyle{1\over 2}}}

\def \vp {\varphi }

\def \t {\theta}
\def \td {\tilde }

\def \ci {\cite}
\def \la {\label}
\def \sm {sigma-model }
\def \foot {\footnote }

\def \inv {^{-1}}
\def \ov {\over }
\def \four{{\textstyle{1\over 4}}}

\def \foot{\footnote}
\def\be{\begin{equation}}
\def\ee{\end{equation}}
\def\bea{\begin{eqnarray}}
\def\eea{\end{eqnarray}}

\begin{document}
\draft
%\date{\today}
\preprint{\vbox{\baselineskip=12pt
\rightline{UPR-0818-T}
\vskip0.2truecm
\rightline{hep-th/9810142}
}}
\title {Microscopics of Rotating Black Holes: \\ Entropy and
Greybody Factors~\footnote{Compilation of talks presented at:
PASCOS'98 (March 22-27, 1998, Boston), Strings'98 
(June 22-27, 1998, Santa Barbara), and  
32nd International Symposium Ahrenshoop
  (September 1-5, 1998, Buckow, Germany).}} 
\author{ Mirjam Cveti\v c{}
}
\address{${}$Department of Physics and Astronomy \\ 
University of Pennsylvania, Philadelphia PA 19104-6396, USA}
\maketitle
\begin{abstract}
We review  the status of  microscopic counting of 
 rotating black hole degrees of freedom. We present
 two complementary approaches
which both utilize the near-horizon geometry and  precisely reproduce the
Bekenstein Hawking entropy for near-extreme rotating 
black holes in D=4 and D=5. The first one, proposed by  Strominger, 
is applicable for the Ramond-Ramond sector and  relies on the
correspondence between the
near-horizon geometry   and the conformal field theory  on its
boundary.  The second one (somewhat more heuristic),
employs the  conformal sigma-model in the Neveu-Schwarz-Neveu-Schwarz 
sector  and accounts for the black hole microstates by counting
the small scale
oscillations of the dyonic string there.
 We also present the 
 wave equation for the 
 minimally coupled  massive scalars for such rotating black hole
backgrounds,  and  discuss its implication for the greybody factors.  The
results are
 illustrated for the prototype  D=5 rotating black hole.
\end{abstract}

%%%%%%%%%%%%%%%%%%%%%%%%%%%%%%%%%%%%%%%%%%%%%%%%%%%%%%%%%%%%
\section{Introduction}
%%%%%%%%%%%%%%%%%%%%%%%%%%%%%%%%%%%%%%%%%%%%%%%%%%%%%%%%%%%%
One of  objectives of the string theory
is to shed light on physics of strong gravitational fields, such as
 issues of  black hole information loss and  related 
issues of  black hole microscopics.  
 In particular,  important  light has been shed on the latter subject of
the  black hole
microscopics.
 
 In a pioneering 
proposal~\cite{sen95} Sen  identified  the microstates of extreme (BPS-saturated) 
electrically charged black
holes  with perturbative excitations of string theory. The proposal was put on a
firmer footing \ci{cmp,harv} by 
interpreting   the  extreme electric 
 black hole states as oscillating modes of 
an underlying macroscopic (fundamental) string.
However,  only after the discovery of { non-perturbative} 
  multi-charge  black holes \cite{kall,dyon} with   { finite}
Bekenstein-Hawking  entropy,   attempts  to  establish  a  
quantitative agreement between the microscopic and macroscopic 
entropy  in string theory became feasible. 
Such solutions were originally  specified by
%as solutions of effective  four-dimensional   toroidally compactified 
%string theory  with  
charges from the  Neveu-Schwarz--Neveu-Schwarz (NS-NS)
sector~\cite{dyon,us1}.  Their microscopic features 
 were captured by string 
theory  in  {\it curved space-time geometry  of the 
near-horizon region}  which is that of 
an  $SL(2,{R}) \times SU(2)$
Wess-Zumino-Witten (WZW) model~\cite{lowe94,us2}.
In particular,  the small-scale  string oscillations~\cite{LW,us2}  were shown 
to reproduce
~\cite{us2,me1,me2} the extreme black 
hole entropy directly from the near-horizon geometry.

%%%%%%%%%%%%%%%%%%%%%%%%%%%%%%%%%%%%%%%%%%%%%%%%%%%%%%%%%%%%%%%%%%
These  developments   were overshadowed  by  the advent of 
D-branes -- nonperturbative objects in string theory with Ramond-Ramond (R-R)
charges~\cite{polch95}.
(Black holes with    NS-NS  charges can be mapped,  by 
duality symmetry,  onto black holes  with   R-R charges  which  
 have 
  the same  space-time metric and thus Bekenstein-Hawking entropy.) The
higher-dimensional 
interpretation  of these black holes  in terms of 
intersecting D-branes  lead to a counting of black hole 
quantum states that agrees precisely with the Bekenstein-Hawking entropy
for
 both the extreme ~\cite{strom96a} and  near-extreme 
\ci{calmal,horstr}  static, as well as rotating ~\ci{ell,all} black holes.

Recently, the implications of   the near-horizon geometry for microscopics
of black holes  was resurrected.  Strominger~\cite{strom} (see also
~\cite{dublinbtz,SS}) has given an alternative derivation of the 
Bekenstein-Hawking  entropy, by again employing  the near-horizon
geometry:
the central observation is that, when embedded in a higher dimensional 
space, the near-horizon geometry contains Ba\~ nados-Teitelboim-Zanelli black
hole (BTZ)~\cite{btz} space-time   which is locally   that of the
three-dimensional 
anti de Sitter space-time ($AdS_3$), whose quantum states are
determined  by a two-dimensional conformal field theory (CFT)  at the
asymptotic boundary 
\ci{adsc}.
The counting of states in this CFT  is  then used to  reproduce 
the Bekenstein-Hawking entropy. 

In the following we review  the microscopics of  rotating black
 holes  within  the two 
complementary approaches, both  utilizing  the  black hole near-horizon
region. 
The first one employs the AdS$_3$/CFT  correspondence~\cite{malda}; the
method  proposed by  Strominger is applied  to  
near-extreme  rotating black holes (with rotation fully
 included)~\cite{cl98a,cl98b}, thus
 generalizing the results for static near-extreme black holes in D=5
 \cite{strom} and
 D=4 \cite{bl98}, respectively.

 The second approach provides  a  counting of 
microstates in the NS-NS sector by relying   
 on the  correspondence  (matching)
between the solitonic  and fundamental string states
 \ci{us2,me1,me2,cvts}. The microstates are counted  directly  as 
string states at the horizon
 and not at the asymptotic boundary. 
Recent
analysis~\cite{cvts} 
  generalizes the   earlier
~\cite{us2,me1,me2}
counting of  microstates for BPS-saturated
 black-holes 
 to the case 
of 
{ non-extreme} black holes. 
The  string sigma-model 
representing a  non-extreme black hole with 
NS-NS charges   is still equivalent to  a (version of)
$SL(2,R)\times SU(2)$ 
WZW  model~\footnote{ A connection of BTZ \cite{btz}  space-time to $SL(2)$ WZW
Lagrangian was explored in \cite{hoor,kalop}.}  Thus  the 
 near-horizon region   remains to be  effectively described 
by the 
free fundamental string whose tension is again rescaled by the 
magnetic charges.    However,  now   the black-hole 
  microstates   are identified  not only with the 
 left-moving, but also {  right-moving}   superconformal  string 
oscillations,
 which  can be  of the same order of magnitude. 

The results are  reviewed for the prototype example of the generating solution
for general D=5 rotating black holes of toroidally compactified  string theory.
(The discussion for D=4 rotating black holes is analogous.) The rest of
the contribution is organized in the following way. In Section II
the black hole parameterization is given  and the  
Bekenstein-Hawking entropy is displayed.  In Section III the near-horizon
geometry
of D=6 rotating dyonic string (obtained by lifting the D=5 black hole
solution) in 
the decoupling limit is discussed and 
the counting  following ~\cite{strom}
is presented. The  minimally coupled massive scalar wave
equation and its implication for the greybody
 factors is discussed in Section IV. In Section V the conformal sigma model,
specified by the near-horizon target space  
of the D=6 rotating string in the NS-NS
sector is addressed and counting of microstates in this sector is presented.
\section{Five-dimensional Rotating Black Hole}
The metric for the classical solution for the  prototype black hole, which correspond to the 
generating solution for the most general D=5 rotating black hole in $N=4$ or
$N=8$ supergravity in five dimensions was given explicitly in ~\cite{cy96a}. 
It is specified by
 $M$, two angular momenta 
$J_{L,R}$, and three independent $U(1)$ charges $Q_i$. 
It is convenient to represent these physical parameters in the
parametric form:
\bea
M &=& m\sum_{i=0}^2 \cosh 2\delta_i~, \\
Q_i &=& m\sinh 2\delta_i~~~;~~i=0,1,2~,\\
J_{L,R} &=& m(l_1\mp l_2)(\prod_{i=0}^2 \cosh\delta_i \pm
\prod_{i=0}^2 \sinh\delta_i)~.
\label{eq:param}
\eea
We work in Planck units where the gravitational coupling 
constant in five dimensions
is $G_5={\pi\over 4}$.

The  Bekenstein-Hawking  entropy  is of the form~\cite{cy96b}:
\bea
S\equiv {A_5\over 4G_5} &=& 
2\pi m \left[ (\prod^2_{i=0}\cosh\delta_i+\prod^2_{i=0}\sinh\delta_i)
(r_++r_-)+\right.
\nonumber \\
&+&
\left. (\prod^2_{i=0}\cosh\delta_i-\prod^2_{i=0}\sinh\delta_i)
(r_+-r_-)\right]~,
\label{eq:macroent}
\eea
where
 \be
r_{\pm}= {1\over 2}\bigg[ \sqrt{2m-(l_1-l_2)^2}\pm 
\sqrt{2m-(l_1+l_2)^2}\bigg]\ ,
\la{rrrp}
\ee
is the chosen value of the radial coordinate  at the outer and inner horizon, respectively.
The two charges $Q_1$ and $Q_2$ appear symmetrically  and can be 
interpreted as
those of a fundamental 
string (FS)  and a NS5-brane and $Q_0$ a momentum along the common direction.
  By duality  $Q_1$ and $Q_2$ charges  
can be interpreted as  those of D1- and D5-brane, respectively. 
D-branes are accessible to a weakly coupled microscopic
description  addressed in the following  Section, while the NS-NS sector
solution has a conformal sigma-model interpetation addressed in Section V.

\section{D=6 Rotating String in Decoupling Limit}
The  above rotating black hole can be lifted 
 to $D=6$ and has an interpretation as a
 dyonic rotating  black string.  Its  Einstein-frame  metric  was 
 given explicitly in \cite{cl98a}. 
Its  internal structure  (and  that of the corresponding black hole)
  can be  
described accurately by a field theory that couples weakly to the 
surrounding space. A precise definition of the decoupling limit 
is given by taking~\cite{malda}:
\be
l_s\rightarrow 0~~;~
r, m, l_{1,2}\rightarrow 0~~;~
\delta_{1,2}\rightarrow\infty~,
\label{eq:decoup}
\ee
where the string length $l_s=\sqrt{\alpha^\prime}$, so that:
\be
rl_s^{-2}~~~;~
ml_s^{-4}~~~;~
l_{1,2} l_s^{-2}~~~;~
Q_{1,2} l_s^{-2}= ml_s^{-2}\sinh 2\delta_{1,2}~,
\ee
remain fixed. The decoupling limit is a near horizon approximation, because
$r\rightarrow 0$, while the ``dilute gas'' conditions
$\delta_{1,2}\rightarrow\infty$ imply that the black hole is necessarily 
near extremal~\cite{greybody}.

The metric of D=6  rotating dyonic string simplifies dramatically in the limit specified
by eq.~\ref{eq:decoup}. 
The near horizon geometry is:
\bea
ds^2_{6} &=&  
{r^2\over f_D\lambda^2}[-(1-{2mf_D\over r^2}) d{\tilde t}^2 + 
d{\tilde y}^2 ]
+ {\lambda^2 r^2 \over (r^2+l_1^2)(r^2+l_2^2)-2mr^2}dr^2 - 
\non\\
&-&2 (l_2\cos^2\theta d\psi + l_1 \sin^2\theta d\phi)d{\tilde t}
-2 (l_1\cos^2\theta d\psi + l_2 \sin^2\theta d\phi)d{\tilde y} + 
\non \\
&+& \lambda^2 (d\theta^2 + \sin^2\theta d\phi^2 + \cos^2\theta d\psi^2)~,
\label{eq:nearmetric} \eea
where
\be
f_D^{-1}= 1+
{{l_1^2\cos^2\theta}\over r^2}+{{l_2^2\cos^2\theta}\over r^2}~,\label{eq:fD}
\ee
and
\bea
{\tilde t} &=& \cosh\delta_0 t - \sinh\delta_0 y~, 
\label{eq:ttilde}
\\
{\tilde y} &=& \cosh\delta_0 y - \sinh\delta_0 t~,
\label{eq:ytilde}
\eea
are the boosted coordinates along the string direction.
The characteristic length scale  $\lambda$ is defined as
$\lambda\equiv (Q_1 Q_2)^{1\over 4}$.

Introducing the shift in the angular variables:
\bea
{\tilde\psi} &=& \psi - \lambda^{-2} (l_2 d{\tilde t}+l_1 {\tilde y})~,
\label{eq:psitilde}
 \\
{\tilde\phi} &=& \phi - \lambda^{-2} (l_1 d{\tilde t}+l_2 {\tilde y})~,
\label{eq:phitilde}
\eea
the metric becomes:
\bea
ds^2_{6} &=&  
-{(r^2+l_1^2)(r^2+l_2^2)-2mr^2\over \lambda^2 r^2 }d{\tilde t}^2 + 
{r^2\over\lambda^2} (d{\tilde y} - {l_1 l_2\over r^2} d{\tilde t} )^2 + \\
&+& 
{\lambda^2 r^2 \over (r^2+l_1^2)(r^2+l_2^2)-2mr^2}dr^2 
+\lambda^2 [d\theta^2 + \sin^2\theta d{\tilde\phi}^2 + \cos^2\theta 
d{\tilde\psi}^2 ]~.
\nonumber 
\eea
In this form it is apparent that the geometry is a direct product
of two three-dimensional spaces. The angular space is a sphere $S^3$ with 
radius $\lambda$, and the geometry with coordinates
$(\tilde{t},\tilde{y},r)$ 
is a BTZ black hole in an effective $2+1$ dimensional theory with 
cosmological constant $\Lambda = - \lambda^2$. 
The metric can be written in the standard BTZ 
form~\cite{btz} with  the effective 
three dimensional mass $M_3$ and angular momentum $J_3$:  
\bea
M_3 &=& 
{R^2_y\over \lambda^4} [ (2m-l^2_1 - l^2_2)\cosh 2\delta_0 +
2l_1 l_2\sinh 2\delta_0 ]~,\\
8G_3 J_3 &=& {R^2_y\over\lambda^3} 
[(2m-l^2_1 - l^2_2)\sinh 2\delta_0 +2l_1 l_2\cosh 2\delta_0 ]~.
\eea
The radius of the compact dimension is denoted by $R_y$.

Thus, the  local space-time remains   that of $AdS_3\times S^3$  (the same as for the
static dyonic string) with the role of  angular
momenta encoded in the global space-time structure!

\subsection{Counting of States}

We now summarize the  counting of  the black hole microstates, 
following~\cite{strom}.
The effective gravitational coupling in three dimensions, $G_3$, can be 
related to the gravitational coupling in five dimensions, $G_5$
($={\pi\over 4}
$), by 
comparing two different dimensional reductions from six dimensions, as 
in~\cite{strom,bl98}. It is:
\be
{1\over G_3} = {1\over G_5}~{A_3\over 2\pi R_y}~,
\ee
where $A_3 = 2\pi^2\lambda^3$ is the area of the $S^3$. This result
is independent of the rotational parameters because the effective 
cosmological constant depends only on the charges of the branes.

The isometry group of the asymptotic $AdS_3$ induces a conformal 
field theory on the boundary at the conformal infinity of the
BTZ black hole. Its central charge is given in terms of the
cosmological constant~\cite{adsc}:
\be
c = {3\lambda\over 2G_3} = 6~{Q_1 Q_2\over R_y}~.
\label{eq:ccharge}
\ee
Note that the central charge is also independent of angular momentum. This
suggests that the rotating black holes can be interpreted as states 
in the same conformal field theory that describes the nonrotating 
black holes.

The relation between the symmetry generators of the induced conformal 
symmetry, and the effective mass and angular momentum are:
\bea
M_3 &=& {8G_3\over\lambda}(L_0 + {\bar L}_0)~,\\
J_3 &=& L_0 - {\bar L}_0~,
\eea
where the eigenvalues of the
operators $L_0$ and ${\bar L}_0$ are the conformal dimensions
$N_L$ and $N_R$, respectively. Then Cardy's formula~\cite{cardy} for the
statistical entropy:
%~\cite{cardy}:
\be
S = 2\pi\left(\sqrt {cN_L\over 6}+\sqrt {cN_R\over 6}\right)~,
\label{eq:cardy}
\ee
gives the microscopic entropy:
\bea
S_{stat}&=& {\pi\over 4G_3}
\left[
\sqrt{\lambda(\lambda M_3+8G_3 J_3)}+\sqrt{\lambda(\lambda M_3-
8G_3 J_3)}\right] \non\\
 &=& \pi\sqrt{Q_1 Q_2}
\left[(r_++r_-)~e^{\delta_0}
+(r_+-r_-)~e^{-\delta_0}\right]~,
\label{eq:microent}
\eea
which precisely matches  macroscopic entropy eq.~\ref{eq:macroent} 
in the limit $\delta_{1,2}\gg 1$.
Thus the microscopic entropy eq.~\ref{eq:microent} precisely
reproduces 
in the decoupling limit eq.~\ref{eq:decoup} where the microscopic calculation 
applies. The range of parameters that are considered here is 
as general as the previous D-brane results~\cite{ell,all}.

\section{Black Hole Perturbations}
\label{sec:waveeq}
The isometry group $SO(2,2)\simeq SL(2,R)_L\times SL(2,R)_R$ of $AdS_3$ 
can be exploited in several ways. The computation of the entropy relies 
on the fact that the BTZ black hole is {\it asymptotically} $AdS_3$, so 
that a conformal field theory is induced at the boundary
at infinity. However, 
the BTZ geometry is in fact {\it locally} $AdS_3$. This has
important consequences for the spectrum of black hole
perturbations~\cite{strom98a,martinecads3,sezginads3,larsen,deboer,gks,martinecb},
and for the dynamics encoded in the greybody factors. Here we discuss some of 
these issues with an emphasis on the effects of rotation.

An important consequence of the local $AdS_3\times S^3$ form of the
metric is that the spectrum of black hole perturbations is organized 
into multiplets of the superconformal algebra. This allows a complete 
classification of all perturbations, as carried out in~\cite{sezginads3}. 
The spectrum of perturbations follows from local properties of $AdS_3$,
and so it is identical for the entire class of black holes considered here.

The perturbations are naturally interpreted
as test fields that interact with the black hole background. The wave 
function of the perturbations then gives the greybody factor, 
expressing the form factor of the Hawking radiation as a  function of 
particle quantum numbers, such as energy and spin, and
the black hole 
parameters~\cite{greybody,cgkt,greybody2,cl97a,cl97b,gubser,hosomichi,cl97d}.
 The greybody factors provide 
a semiclassical testing ground for dynamical properties. In the special 
case of minimally coupled scalars in the S-wave they agree precisely with 
calculations in string theory~\cite{mathur}. The microscopic processes
responsible for other modes can be modelled in terms of an 
effective string theory with dynamics that reproduces the black hole 
greybody factors qualitatively~\cite{cgkt,greybody2}. 

\subsection{Minimally Coupled Massive Scalar Wave Equation}
For the sake of concreteness we  discuss the  case of a minimally 
coupled  massive scalar field. By using the  original coordinates $t,y$ and
the radial variable $x$ defined as 
\be
x\equiv
{r^2 - {1\over 2}(r^2_+ + r^2_-)\over r^2_+ - r^2_-},
\label{eq:xdef}
\ee
the  near-horizon metric eq. \ref{eq:nearmetric} is of the form:
\bea
ds^2_6 &=& \lambda^{-2} 
\left[- (x-{1\over 2})~ (r_+ d{\tilde t} - r_- d{\tilde y})^2
+ (x+{1\over 2})~ (r_+ d{\tilde y} - r_- d{\tilde t})^2 \right] + 
\label{eq:xwaveeq} \\
&+& \lambda^2 \left[{1\over 4x^2 - 1} dx^2 + 
d\theta^2 + \cos^2 \theta 
d{\tilde\psi}^2 + \sin^2 \theta d{\tilde\phi}^2 \right]~,
\nonumber
\eea
where $({\tilde t}, {\tilde y})$ 
and  $({\tilde\psi}, {\tilde\phi})$
are defined in eqs.~\ref{eq:ttilde}--\ref{eq:ytilde} and 
eqs.~\ref{eq:psitilde}--\ref{eq:phitilde}, respectively.
Inserting the  following  Ansatz for the wave function
~\footnote{
The fact that there is a separation of  variables for this 
rotating background  with the massive minimally coupled scalars 
is nontrivial. (The same observation was made for  massless minimally coupled 
scalars in the five dimensional black hole geometry~\cite{cl97a}.) It indicates
that there may be additional conserved Killing-Stackel-type  tensors, 
analogous to the  case of  Kerr black holes in four dimensions.
(For some discussion and references see~\cite{susysky,cl97c}.)}:
\be
\Phi= \Phi_0(r)~\chi(\theta)~e^{-i\omega_R (t+y)-i\omega_L (t-y)
+im_R(\phi+\psi)+im_L(\phi-\psi)}~,
\label{eq:ansatz}
\ee 
into the Klein-Gordon equation:
\be
{1\over\sqrt{-g}}\partial_\mu (\sqrt{-g}g^{\mu\nu}\partial_\nu \Phi) = 
\mu^2 ~,
\label{eq:kg}
\ee
one finds the following form of the radial wave equation~\cite{cl98a}:
\bea
&\big[&{\partial\over\partial x}(4x^2-1){\partial\over\partial x}
+{1\over x-{1\over 2}}
({\beta^R\omega_R+\beta^L\omega_L-m_R\beta_H\Omega^R
-m_L\beta_H\Omega^L \over 2\pi} )^2 \non \\
&-& {1\over x+{1\over 2}}
({\beta^R\omega_R-\beta^L\omega_L-m_R\beta_H\Omega^R
+m_L\beta_H\Omega^L \over 2\pi} )^2 \big]\Phi_0 = 
(\Lambda+\lambda^2\mu^2)\Phi_0~.
\label{eq:geneq}
\eea
Here the eigenvalues of the angular 
Laplacian:
\be
{\hat\Lambda} =
- {1\over \sin 2\theta}{\partial\over\partial\theta}
\sin 2\theta{\partial\over\partial\theta}
-{1\over \sin^2 \theta}{\partial^2\over\partial\phi^2}
-{1\over \cos^2 \theta}{\partial^2\over\partial\psi^2}~,
\label{eq:flatlap}
\ee
are denoted $\Lambda$ and take the form $\Lambda=l(l+2)$ 
where $l=0,1,\cdots$.
In eq. \ref{eq:geneq}
 \be
\beta^{L,R} = {2\pi\lambda^2 e^{\mp\delta_0}\over
\sqrt{2m-(l_1\mp l_2)^2}} ~, \ \  
\beta_H\Omega^{L,R} ={2\pi (l_1\mp l_2)\over\sqrt{2m-(l_1\mp l_2)^2}}~.
\label{eq:omega}
\ee
$\beta^{L,R}$ have an interpretation of potentials  
conjugate to the left- and right-moving energy 
of the 2d-CFT  (with $\beta_H=\textstyle{1\over 2} (\beta_L+\beta_R)$
corresponding to the inverse of the Hawking temperature);
 $\beta_H\Omega^{L,R}$ are conjugate to
the two  angular momenta $J_{L,R}$, respectively.

The wave equation eq.~\ref{eq:geneq} agrees 
with the near horizon limit of the general one given in~\cite{cl97a},
except that there the {Ansatz} for the wave function did not allow 
dependence on the compact coordinate $y$~\footnote{An equation that
applies in the general nonextremal case and includes the 
dependence on the compact coordinate $y$ can be obtained from the master 
equation given in~\cite{cl97a}, 
by exploiting boost invariance in the $y$ dimension. In ~\cite{cl97a} it was also
observed that the near-horizon region of the  radial wave equation respects 
the $SL(2,R)_L\times SL(2,R)_R$
symmetry leading to a  conjecture that this symmetry has  a 2-d CFT origin. 
However, a connection of this CFT to  the $AdS_3$ was not recognized, since  
in the five-dimensional
context the dependence on the $y$
direction was suppressed.}. The present 
generalization gives an even more symmetric result.

It follows immediately  from the form of the wave equation eq. \ref{eq:geneq} 
that its 
solutions depend on $m_{L,R}$ and $\beta_H\Omega^{L,R}$ only through the
prescriptions:
\be
\beta^{L,R}\omega_{L,R}\rightarrow 
\beta^{L,R}\omega_{L,R}- m_{L,R}\beta_H\Omega^{L,R}~.
\label{eq:blrshifts}
\ee
This rule gives the exact wave functions in the rotating background,
when the nonrotating ones are known. (It turns out to be valid for all fields in the 
near-horizon region, without regard to the details of their couplings.) 
  Note also that   a mass $\mu$ is included in the Klein-Gordon
equation, eq.~\ref{eq:kg}, and its effect can be absorbed 
in the conformal dimension. The solution to the
near horizon wave equation is a hypergeometric function found
in~\cite{greybody2,gpartial,mpartial} (present notation is used 
in~\cite{cl97a}).

\section{Conformal sigma model for rotating black holes}
We now turn to the study of black hole microscopics in the NS-NS sector.
We  demonstrate~\cite{cvts} that   for a large NS5-brane   and 
fundmental string 
the  black hole lifted to D=6 is 
 described by a marginal integrable  perturbation 
of the $SL(2,R)\times SU(2)$ WZW model generated  by  the 
left- {\it and}  right-moving  Cartan 
chiral  $SU(2)$ currents 
whose  perturbation parameters are
proportional to the two angular momentum components.

In NS-NS sector the D=6 rotating dyonic   string solution is described
  by the three NS-NS charges: the
NS5-brane charge $Q_1$,  the fundamental string (FS) charge $
Q_2$  and the 
string  momentum  $Q_0$. Its Einstein frame metric is the same as
that  specified by R-R-charges.
The string  \sm   corresponding to  this 
background  has a complicated form, however it simplifies significantly in
the
near-horizon region.
 In this limit, it takes  
   the following form~\cite{cvts}~\footnote{For the sake of making the
notations of
   Sections III,  and V compatible,  the convention here differs
    from that
   of \cite{cvts} in the following way: 
%$(\del,\bd)_{here}=(\bd, \del)_{there}$,
   $(Q_0,Q_1,Q_2)_{here}=({\tilde
   Q},P,Q)_{there}$,  $(J,{\bar J})_{here}= 2({ J},{\bar J})_{there}$,
   and $e^{{\delta_0}}_{\ \ here}=e^{-\delta_{\tilde Q}}_{\ \ \ \ there}$. }: 
\bea
(L)_{\delta_{1,2}\gg 1}
&=& 
{Q_2\inv } \left[(r^2f^{-1}_D-m) \del U\bd V 
 +\textstyle{1\over 2} m e^{2\delta_{0}}\del U  \bd U + 
 \textstyle{1\over 2}me^{-2\delta_{0}} \del V \bd V\right] +
\non\\
&-& \sqrt{Q_1\over Q_2}\left[(l_1-l_2) 
e^{\delta_{0}}\ \del U{\bar J}_3  
+(l_1+l_2)
e^{-\delta_{0}}\ J_3\bd V\right]  
\non\\
&+&   Q_1\left[{{r^2 \del r\bd r}\over {(r^2-r_+^2)(r^2-r_-^2)}} 
+   L_{SU(2)}\right] \ ,  \la{def}
\eea
 with  the constant dilaton $e^{2\Phi}={Q_1\over Q_2}$. Here  the light-cone
 string coordinates are defined as: $(U,V)=\mp t+y$, while 
\be
L_{SU(2)}=
 \del \t \bd \t  +
 \sin^2\theta\del \vp \bd \vp   +  \cos^2\theta\del \psi \bd \psi
  , \ee
is the $SU(2)$ WZW Lagrangian with the  $SU(2)$ chiral
Cartan currents:
\be
J_3= \sin^2\t \del \vp \  + \cos^2 \t\ \del \psi\ , 
\ \ \ \ \ \ \ \ \
\bar J_3= -\sin^2\t \bd \psi \ +  \cos^2 \t\ \bd \vp\ .
\la{carr}
\ee
It is apparent  that the Lagrangian eq. \ref{def} can be interpreted as a
chiral marginal perurbation of a  \wzw   WZW model by  the  $SU(2)$ 
Cartan currents eq. \ref{carr}.
This can be confirmed  by noting that 
 after the  coordinate transformation~\cite{cl98a}
\bea
{\vp}&\to&  \vp - \textstyle{1\over 2} (Q_1Q_2)^{-{1\over 2}}
\left[-(l_1-l_2)e^{\delta_{0}}
U  +(l_1+l_2)e^{-\delta_0}V\right]~, \non \\ 
{  \psi}&\to&   \psi - \textstyle{1\over 2}(Q_1Q_2)^{-{1\over 2}}
\left[(l_1-l_2)e^{\delta_{0}}
U  +(l_1+l_2)e^{-\delta_0}V\right]~,
\la{fred}\eea
the Lagrangian  eq. \ref{def}  assumes the form:
\bea
(L)_{\delta_{1,2}\gg 1}&=&{Q_2\inv } [\left(r^2- \ha(
r_+^2+r_-^2)\right) \del 
U\bd V 
+\textstyle{1\over 4} (r_++r_-)^2e^{2\delta_{0}} \del U  \bd
U \non\\ &+&\textstyle{1\over 4} (r_+-r_-)^2e^{-2\delta_{0}} \del V \bd V]
+  Q_1\big[{{r^2 \del r\bd r}\over {(r^2-r_+^2)(r^2-r_-^2)}} 
+L_{SU(2)}\big]\ ,
\label{eq:final}  \eea
 which is  precisely the \wzw  WZW Lagrangian at  level $\kappa=Q_1$
 ~\footnote{
 Note that  transformations:
${\cosh {z\over 2}}= {\sqrt{r^2- r_-^2\over {r_+^2-r_-^2}}}$, 
$u= {1\over {2\sqrt{Q_1Q_2}}} (r_++r_-)e^{\delta_{0}}U $ and \newline
$v= {1\over {2\sqrt{Q_1Q_2}}} (r_+-r_-)e^{-\delta_{0}} V$
render  
 $L_{SL(2)}$ part of the Lagrangian  eq. \ref{eq:final} in  the canonical form:
$L_{SL(2)} =  \k (2 \cosh z \ \del u \bd v + 
     \del u  \bd u +   \del v  \bd v  +    \del z \bd z )$ where
 $\k=Q_1$.}! 
%%%%%%%%%%%%%%%%%%%%%%%%%%%%%%%%%%%%%%%%%%%%%%%%%%%%%%%%%%%%%%%%%%%%%%%%%%%%%%%%%%%%%%%%%%%%%
\subsection{Statistical  entropy}
\label{entropy}
%%%%%%%%%%%%%%%%%%%%%%%%%%%%%%%%%%%%%%%%%%%%%%%%%%%
The aim will be to  employ the  structure of the \wzw model eq. \ref{eq:final}
to give  a 
microscopic interpretation  of the Bekenstein-Hawking entropy
eq. \ref{eq:macroent}
 in the limit
of large  $Q_1,\ Q_2$ charges. 

The main idea (originated in \cite{us2}, further refined in \cite{me1,me2} and
generalized to  include the right-moving oscillations in \cite{cvts})
 is that in the large charge  limit  the  \wzw WZW model  describes
a fundamental string with a tension  rescaled by the NS5-brane charge 
$Q_1$, 
and whose 
relevant marginal deformations 
are  only those of 
 the four transverse  
directions to the NS5-brane.  
 Since the charges are large, 
the conformal \sm   eq. \ref{eq:final} is weakly coupled, so one is 
 to count  string states
 in nearly-flat space. This implies  that the central $c$
charge  of the ``free'' effective  superstring  that accounts for 
the bosonic and fermionic excitations  in the four transverse
directions  is 
\be
c = 4(1+\textstyle{1\over 2})=6~. 
\la{eq:c}
\ee
In order to relate the oscillation numbers $N_{L,R}$ of the  fundamental
string and thus 
its entropy to the global parameters  of the  black hole 
 one  is to use the ``matching conditions'' 
 analogous to the ones in  \ci{cmp,harv,me2}. Below we describe the counting 
procedure of the small scale oscillations for  the 
   WZW Lagrangian eq. \ref{eq:final}.

Let us  combine the  $z$-direction (related to $r$) of $L_{SL(2)}$ 
 Largangian,  
 and three  angular
 coordinates of   ($\t,\ \vp , \ \psi$)
of $L_{SU(2)}$ Lagrangian  into  four  transverse coordinates  $x^i$.
Since  the  WZW level  $\k=Q_1$ is large,
the interactions in  WZW theory are suppressed
 by $1/\k$, and thus   the count of perturbations should be essentially the 
same as in the theory
of free fields $x^i$. 
The only difference compared to the flat space case is the presence of the 
factor $\k$ in front of the kinetic term of $x^i$ coordinates (cf. eq.
\ref{eq:final}). 
The sigma-model  which represents the  perturbed
version of the 
model eq. \ref{eq:final} by left- as well as right-moving perturbations can be written as
\bea\non
  L&=&Q_2^{-1}[F(x) \del U \bd V  + K_0 \de U \bd U + M_0 \de V  \bd V
\\
&+& \  \A_i (x,U) \de U \bd x^i  +  \AA_i (x,V) \de x^i\bd V] 
+   Q_1 h_{ij}(x) \de x^i \bd x^j \  . 
\la{gene}
\eea 
  Here $(U,V)$ 
correspond to the  string  light-cone coordinates $(U,V)=\mp t+y$
of the  effective  fundamental string. 
Since the level $\k=Q_1 $ is large, we may assume 
that $h_{ij}$ is approximately flat.
%and  also set  $F(x) 
%\approx 1$ (the large charge limit corresponds to 
%in the large $P$ charge  
% the small curvature of  $AdS_3$ space).
  As a result, the perturbations  $\A_i (x,U)$
and $\AA_i (x,V)$ are marginal to the leading order in their strength. Indeed, 
integrating out $U$ ``freezes'' $V$ and thus makes 
$\AA_i $ marginal and vice versa for $\A_i$. 
In general, there is also a constraint on their $x$-dependence,
and  the solution which is relevant in the large charge limit is 
\be\la{soo}
\A_i (x,U) \sim a_i(U)\   , \ \ \ \ \ \  \ 
\AA_i (x,V) \to b_i(V)\ . 
\ee
Integrating out the four transverse fields $x^i$ 
in the large $\k$ limit we find
that $K_0$ and $M_0$ in \rf{gene}  are replaced by
\be
\la{rep}
K(U) = K_0 - (Q_1Q_2)\inv a^2_i(U)~, \ \ \
M(V) = M_0 - (Q_1Q_2)\inv b^2_i(V) ~.
\ee
(In general, there is also a  $\A\AA$-correction to 
the $\del U \bd V$ term.)

The analogue of the level matching condition 
for the free fundamental string \ci{cmp,harv}
which relates  the oscillation level numbers $N_{L} $ and $N_{R}$ 
to the  background 
charges is  as follows:   the coefficients of  both  $\del U \bd U$ and
$\del V \bd V$ terms  should vanish on
 average to allow matching onto a fundamental  string source, 
i.e. 
\be
\overline{ K } =0 \ , \ \ \ \  \ \ \ \ \ \overline{ M } =0 \ , 
\ \ \ \ \ \ \ \   \overline{  f }
  \equiv {1 \ov 2\pi R} \int^{2\pi R}_0 dy \  
f(y) \ ,  \la{kyk}
\ee
where $R$ is the radius of  a compact direction along which the string is wound.
Then,  to the leading order in $1/\kappa$,  
\be
   \overline{ a^2_i}=Q_1Q_2 K_0  =  N_L  \ , \ \ \ \ \ \ \ \ \  
\overline{ b^2_i} = Q_1Q_2M_0= N_R  \ ,   \la{reee}
\ee
%$$ q\equiv{\pi \ov 4 G_N  } \ , $$
where  $N_L$ and $N_R$ are interpreted as the left- and the right-moving 
string oscillation numbers. (The 
proportionality  constant between $N_{L,R}$ and $(K_0,M_0)$
depends  on  the  tension of the  
fundamental string and is equal to 1
 in  units  used (i.e., $G_5={\pi\over 4}$).)

Note that for  the Lagrangian eq. \ref{eq:final}  the  relationship eq.
\ref{reee} implies:
\bea
N_{L} &=&Q_1Q_2 K_0= \four Q_1 Q_2 (r_++r_-)^2
e^{2\delta_{0}}~, \non\\
N_R&=&Q_1Q_2M_0= \four Q_1Q_2 (r_+-r_-)^2e^{-2\delta_{0}}~. 
\label{eq:N}\eea
Inserting the value of the central charge eq. \ref{eq:c} and the oscillator
numbers  $N_{L,R}$  eq. \ref{eq:N} into Cardy's entropy formula 
 eq. \ref{eq:c} one
obtains:
\be
S_{stat}= \pi\sqrt{Q_1 Q_2}
\left[(r_++r_-)~e^{\delta_0}
+(r_+-r_-)~e^{-\delta_0}\right]~,
\ee
which is indeed in agreement with that of eq. \ref{eq:macroent} in the limit if
large charges.
%%%%%%%%%%%%%%%%%%%%%%%%%%%%%%%%%%%%%%%%%%%%%%%%%%%%%%%%%%%%%%
 \section*{Acknowledgments}

I would like to thank F. Larsen and A. Tseytlin for collaboration on the work
presented in this contribution.
The work is supported in part by DOE grant DOE-FG02-95ER40893.

%%%%%%%%%%%%%%%%%%%%%%%%%%%%%%%%%%%%%%%%%%%%%%%%%%%%%%%
%%%%%%%%%%%%%%%%%%%%%%%%%%%%%%%%%%%%%%%%%%%%%%%%%%%%%%%%

% \section*{References}

\end{document}